\documentclass[a4paper,english,reprint,superscriptaddress,longbibliography,aps,prl]{revtex4-1}
\usepackage{lmodern}

\usepackage[T1]{fontenc}
\usepackage[utf8]{inputenc}
\usepackage{babel}
\usepackage{float}
\usepackage{amsmath}
\usepackage{amssymb}
\usepackage{graphicx}
\usepackage[pdfusetitle,
 bookmarks=false,
 breaklinks=false,pdfborder={0 0 0},pdfborderstyle={},backref=false,colorlinks=false]
 {hyperref}

\makeatletter

\pdfpageheight\paperheight
\pdfpagewidth\paperwidth

\usepackage{braket}
\usepackage{balance}

\makeatother

\begin{document}
\title{Observation of the Talbot effect from a surface acoustic wave dynamic
grating}
\author{M. Fisicaro}
\email{fisicaro@physics.leidenuniv.nl}

\affiliation{Huygens-Kamerlingh Onnes Laboratory, Leiden University, P.O. Box 9504,
2300 RA Leiden, The Netherlands}
\author{Y. C. Doedes}
\affiliation{Huygens-Kamerlingh Onnes Laboratory, Leiden University, P.O. Box 9504,
2300 RA Leiden, The Netherlands}
\author{T. A. Steenbergen}
\affiliation{Huygens-Kamerlingh Onnes Laboratory, Leiden University, P.O. Box 9504,
2300 RA Leiden, The Netherlands}
\author{M. P. van Exter}
\affiliation{Huygens-Kamerlingh Onnes Laboratory, Leiden University, P.O. Box 9504,
2300 RA Leiden, The Netherlands}
\author{W. Löffler}
\affiliation{Huygens-Kamerlingh Onnes Laboratory, Leiden University, P.O. Box 9504,
2300 RA Leiden, The Netherlands}
\begin{abstract}
We demonstrate the dynamical Talbot effect caused by optical diffraction
from standing surface acoustic waves (SAWs). The Talbot effect is
a wave interference phenomenon in the Fresnel regime, and we observe
it with a fiber-based scanning optical interferometer on a SAW Fabry-Perot
cavity. By studying the interferometric signal at 1 GHz, we first
discover the existence of an amplitude-modulated term, that can exceed
in magnitude the usual phase-modulated term, enabling a new way of
imaging surface acoustic waves. Secondly, by displacing the acoustic
device from the beam focus we reveal the optical Talbot effect, where
despite the curved wavefronts of the optical field, the conventional
Talbot length appears. As a consequence, the amplitude modulation
vanishes at periodic positions of the acoustic wave relative to the
beam focus.
\end{abstract}
\maketitle
Diffraction of a plane wave by a periodic structure such as a grating
leads to periodic images of the grating at specific distances – therefore
the Talbot effect is also known as lensless imaging or self-imaging
\citep{montgomerySelfImagingObjects1967,patorskiSelfImagingPhenomenon1989,wenTalboteffect2013}.
It was firstly discovered in optics in 1836 by Henry Fox Talbot \citep{talbotLXXVIFacts1836},
but it was Lord Rayleigh who quantitatively explained it in 1881 \citep{rayleighXXVcopying1881}.
Four years later, Lord Rayleigh also predicted the existence of a
particular type of acoustic surface waves, characterized by an elliptical
motion of the surface \citep{rayleighWavesPropagated1885}. He accurately
envisioned that these waves have a significant role in earthquakes
and elastic solids, and they were later named Rayleigh waves. Despite
the Talbot effect has been observed in many different systems \citep{chapmanNearfieldimaging1995,songExperimentalObservation2011,candelasObservationultrasonic2019,bakmanObservationTalbot2019,gaoTalbotEffect2016,zhangNonlinearTalbot2010,rozenmanPeriodicWave2022},
to the best of our knowledge, it has never been observed as a consequence
of optical diffraction from Rayleigh waves.
\begin{figure}[t]
\includegraphics[width=1\columnwidth]{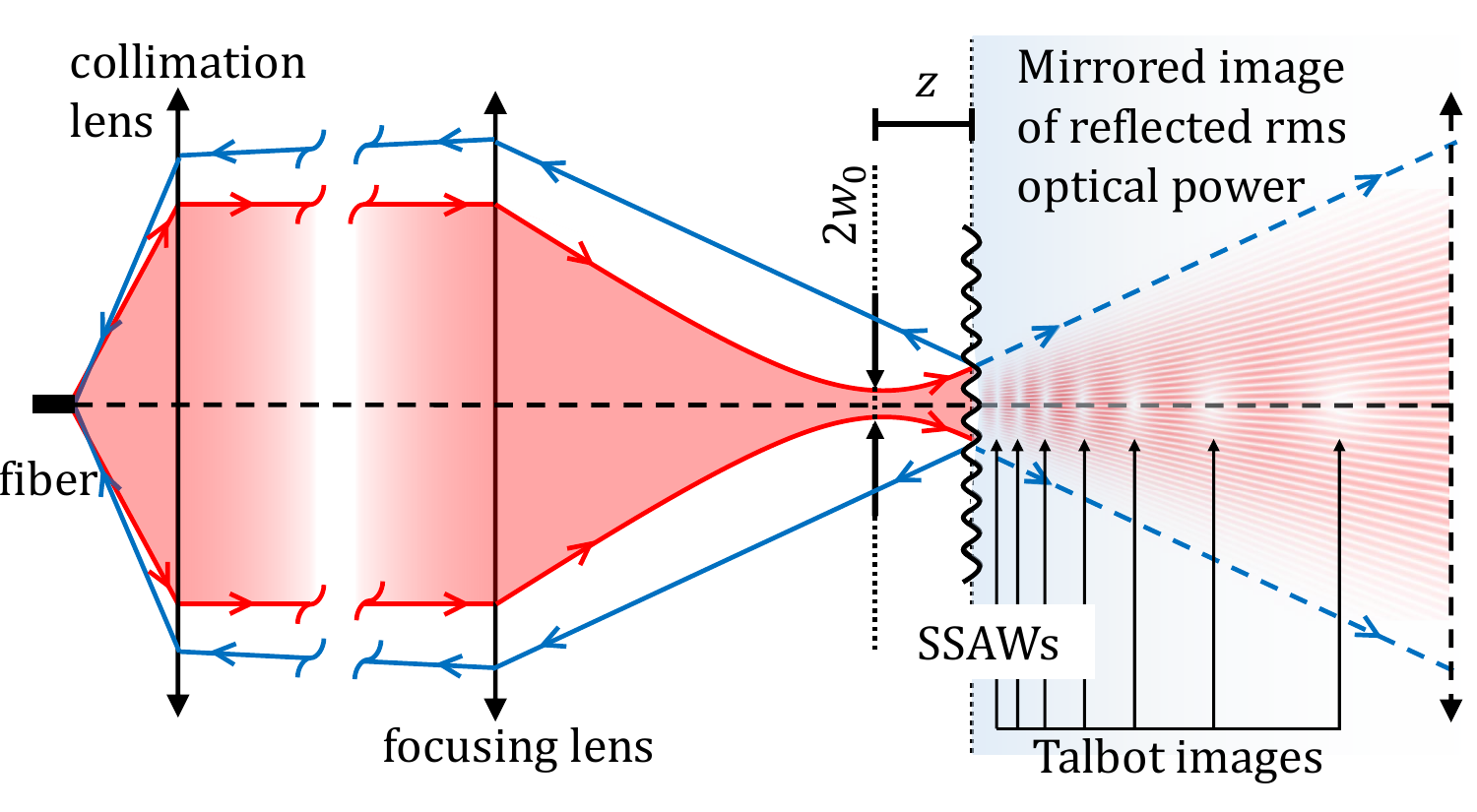}

\caption{\protect\label{fig:Sketch_of_experiment}Sketch of the experiment:
laser light from a single mode fiber is first collimated and then
focused onto a partially reflecting surface on which standing surface
acoustic waves (SSAWs) are excited. The incident light is reflected
back, for clarity we plot here its mirrored image to the right of
the reflecting surface. If the position of the beam waist $z$ is
displaced with respect to the reflecting surface, the light beam illuminates
multiple periods of the oscillating surface grating and the Talbot
effect occurs. The self-images of the field at the grating, produced
by the Talbot effect, are visible as regions of zero power on the
right of the grating, where we have plotted the rms optical power
of the reflected diffracted beam.}
\end{figure}

Surface acoustic waves (SAWs) can be excited using an interdigital
transducer (IDT) \citep{joshiExcitationDetection1969,mamishevInterdigitalsensors2004a,whiteDirectpiezoelectric1965}
on the surface of a piezoelectric substrate at frequencies up to a
few GHz, which for most materials corresponds to acoustic wavelengths
down to the sub-micrometer range \citep{delsing2019surface2019a,yamanouchiSHFrangesurface1988,buyukkose,wangHighperformance2018}.
If confined in an acoustic Fabry-Perot cavity \citep{shaoPhononicBand2019,schuetzUniversalQuantum2015a,msallFocusingSurfaceAcousticWave2020,takasuSurfaceacousticwaveresonators2019a,luschmannSurfaceacoustic2023a,manentiSurfaceacoustic2016a,magnussonSurfaceacoustic2015,mooresCavityQuantum2018}
standing SAWs appear, resulting in an oscillating surface grating,
which can dynamically diffract an optical beam upon reflection by
spatial phase modulation. In our case, we use a fiber-based scanning
optical Michelson interferometer to image the displacement generated
by 1 GHz SAWs. For spatially-resolved interferometric imaging of the
SAWs shown in Fig. \ref{fig:Sketch_of_experiment}, this diffraction
is an in principle unwanted but unavoidable effect since at this frequency,
the acoustical wavelength $\Lambda=2.8\,\mathrm{\mu m}$ and the beam
spot size of the focused laser beam $2w_{0}=2.8\,\mathrm{\mu m}$
(for $\lambda=980\,\mathrm{nm}$ and 0.55 NA) are comparable in size.
In our experiment, we were originally interested in the optical phase
modulation induced by the SAWs, but we found that the interferometric
signal consists of both a phase modulation term and of an amplitude
modulation term that appears due to mode filtering by the single-mode
optical fiber. In this study, we theoretically and experimentally
show that the amplitude modulation term can be significant, enabling
a simpler way of measuring SAWs, and we report the observation of
the Talbot effect from a standing SAW grating, sketched in Fig. \ref{fig:Sketch_of_experiment}.

\begin{figure}[t]
\includegraphics[width=1\columnwidth]{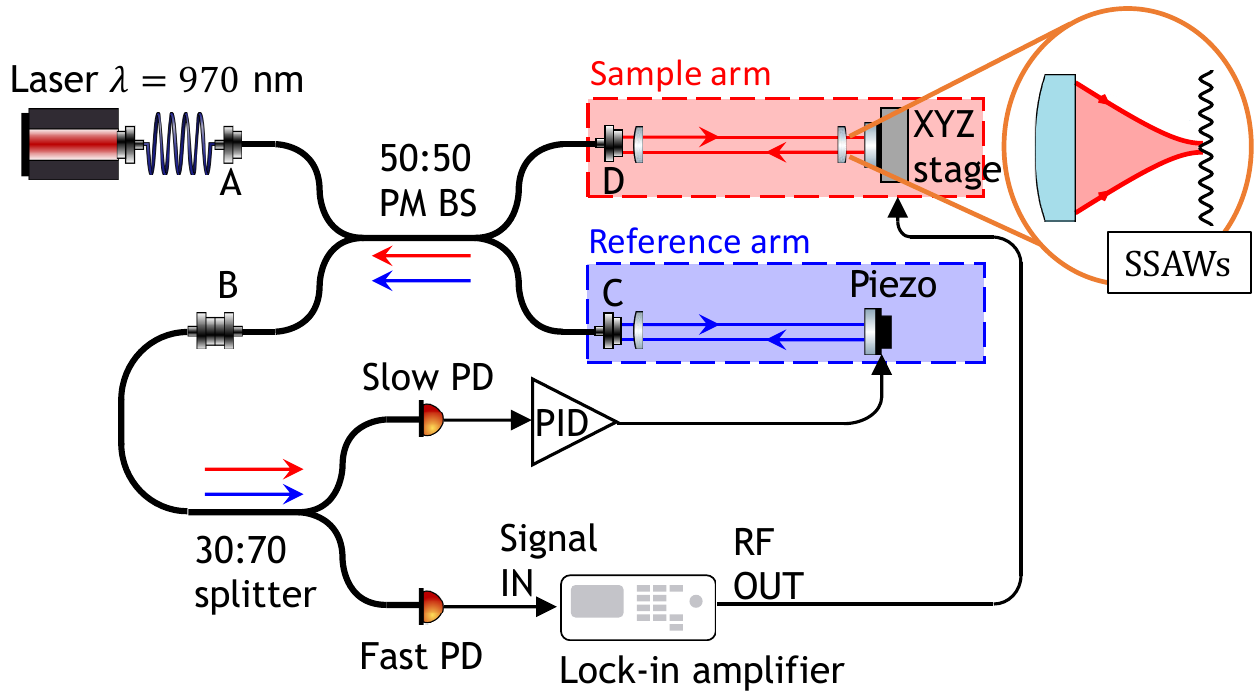}

\caption{Scheme of our experiment, a fiber-based Michelson interferometer using
single-mode polarization-maintaining fibers and splitter. PD: photodiode.
\protect\label{fig:experimental_setup}}
\end{figure}

\begin{figure}[t]
\includegraphics[width=1\columnwidth]{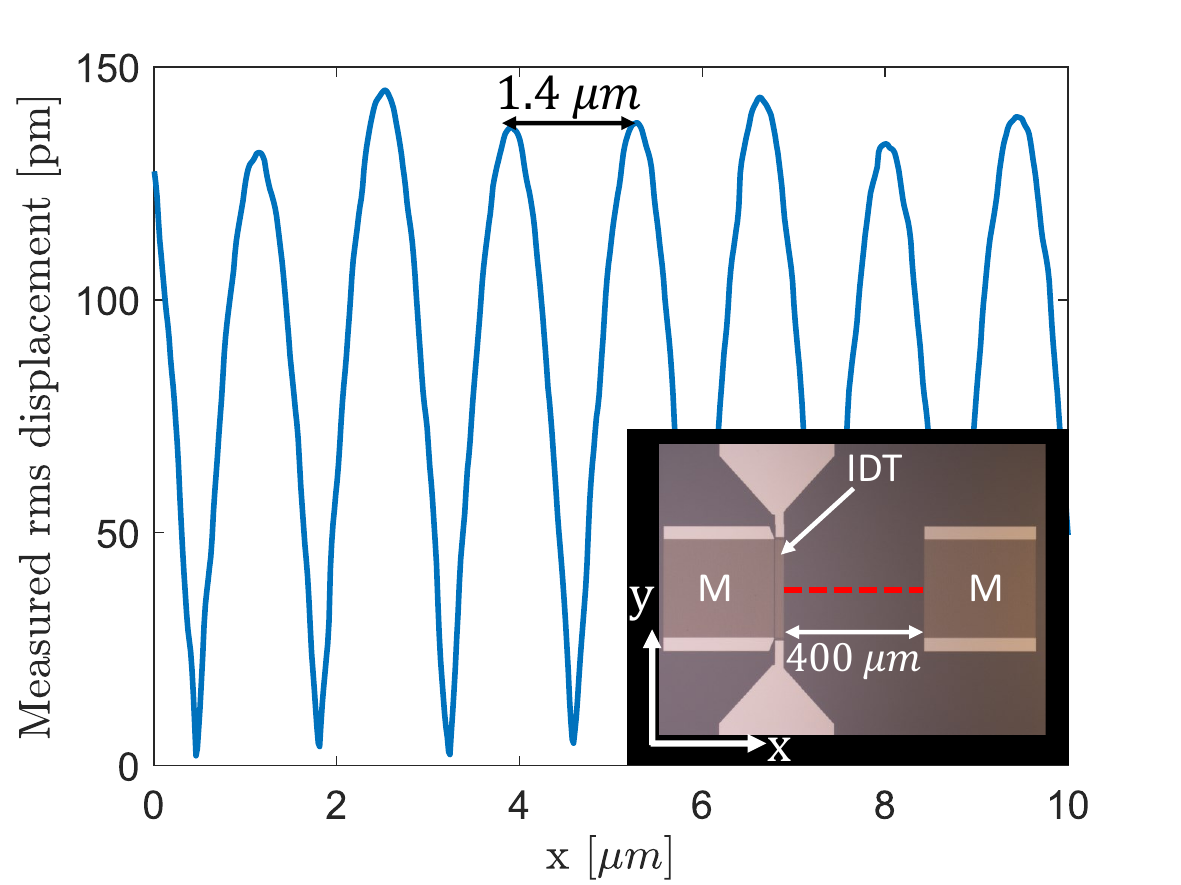}

\caption{Microscope image (inset) of the SAW cavity showing the IDT and the
SAW mirrors (M), and standing wave pattern measured along a short
section of the dashed red line, revealing the expected periodicity
of $\Lambda/2=1.4\,\textrm{\ensuremath{\mu}m}$. The rms displacement
is $\sim140\,\textrm{pm}$ at the antinodes. \protect\label{fig:Standing_waves_and_SAW_cavity}}
\end{figure}

Our experimental setup shown in Fig.~\ref{fig:experimental_setup}
is a Michelson interferometer implemented with a polarization-maintaining
single-mode fiber coupler as the beam splitter (PM BS). Narrow line-width
laser light of wavelength $\lambda=980\,\mathrm{nm}$, after optical
isolation, enters the fiber coupler through port A and is split into
the sample (D) and reference arm (C). In the reference arm, the light
is focused onto a mirror by a single aspheric lens and back-reflected
into the fiber. In the sample arm (D), the light is first collimated
and then strongly focused onto the GaAs-based SAW device with an aspheric
lens with 0.55 NA resulting in $w_{0}=1.4\ \mathrm{\mu m}$, and the
back-reflected light is coupled back into the same single-mode fiber.
Both light fields are combined in the polarization-maintaining splitter,
and detected by a slow and a fast photodiode. The SAW device is a
surface acoustic wave (SAW) Fabry-Perot cavity fabricated on a GaAs
substrate, where SAWs are excited by an interdigital transducer with
10 finger pairs inside the resonator, see inset in Fig. \ref{fig:Standing_waves_and_SAW_cavity}.
The SAW mirrors are implemented as short-circuited metal gratings
with 250 elements each, 50 nm thick. The device is designed for a
frequency of 1 GHz corresponding to a SAW wavelength of $\Lambda=2.8\,\mathrm{\mu m}$.
The signal on the slow photodiode in combination with a PID controller
and a piezo in the reference arm is used to stabilize the interferometer.
The SAW device is mounted on a three-axes nanopositioning stage, where
translation along the optical ($z$) axis allows to adjust the focus
of the laser beam, while the $x$ and $y$ axes allow us to scan the
laser focus over the SAW device, in order to reconstruct an image
of the out-of-plane displacement of the SAW device. Fig. \ref{fig:Standing_waves_and_SAW_cavity}
shows a microscope image of the SAW device, and a cross-section of
the standing SAW displacement if the device is in focus ($z=0$).
However, we have found that also with blocked reference arm, sometimes,
the SAW signal appears very clearly. This originates from a complicated
form of periodic focusing and defocusing of the reflected beam, which
modulates its collection efficiency in the single mode fiber. We call
the signal in the conventional case \emph{interferometric signal},
and if the reference arm is blocked, the \emph{amplitude modulation}.
We will now develop a theoretical model explaining this effect, importantly,
we not only consider the case that the SAW device is in focus, but
at a distance $z$ from focus (Fig. \ref{fig:Sketch_of_experiment}).

Since we use throughout the setup linearly polarized light, it is
sufficient to use scalar fields for the input field $E_{in}$, the
collected field $E_{col}$, and the reference field $E_{ref}$. We
introduce the complex collection ratio
\begin{equation}
\alpha(t)\ =\ \frac{E_{col}(t)}{E_{in}},\label{eq:alpha}
\end{equation}

which is a measure of the amplitude and phase of the reflected light
collected by the single mode fiber, relative to the input light that
enters the sample arm of the interferometer. Once we know $\alpha(t)$,
we can write the total intensity at the fast photodiode as
\begin{align}
I_{tot}\  & =\ |E_{ref}\ +\ \alpha(t)\ E_{in}|^{2}\label{eq:Intensity_interference}\\
 & =\ I_{ref}\ +\ |\alpha(t)|^{2}I_{in}\ +\ \mathrm{Re}[2\alpha(t)E_{in}E_{ref}^{*}],\nonumber 
\end{align}

where Re denotes the real part. We also note that while phase variations
in $\alpha(t)$ are only visible in the double product term of Eq.
\ref{eq:Intensity_interference}, amplitude variations are also visible
in the $|\alpha(t)|^{2}I_{in}$ term and can be observed also with
blocked reference arm when $E_{ref}=0$.

The calculation of the complex collection ratio $\alpha(t)$ is based
on the overlap integral between the Gaussian mode supported by the
fiber and the beam reflected by the SAW device, as a function of the
defocusing, namely the distance $z$ between the SAW device surface
and the beam waist position of the focused Gaussian beam. This overlap
integral can be calculated at any plane, for simplicity we choose
to calculate it on the device surface. 

Due to the presence of standing SAWs along the $x$ axis, we can model
the surface out-of-plane displacement as

\begin{equation}
\Delta d(x,y,t)\ =\ A(t)\ \cos\left[K\left(x-x_{0}\right)\right],\label{eq:SSAWs_displacement}
\end{equation}

where $K=2\pi/\Lambda$ is the SAW wave number, $x_{0}$ is the transverse
distance between the center of the laser focus and the position of
the standing SAW anti-node, and $A(t)=A_{0}\cos(\Omega t)$, is the
time-dependent amplitude of the SAW with angular frequency $\Omega$
and peak displacement $A_{0}$. There is no dependency on $y$ because
the standing SAW are excited only along the $x$ axis. In Eq. \ref{eq:alpha},
we defined the complex collection ratio as a function of time, but
now we compute it also as a function of $z$ and $x_{0}$: $\alpha(t)\to\alpha(z,x_{0},t)$.
From calculations shown in Appendix \ref{sec:complex_collection_ratio}
we obtain:

\begin{align}
\alpha(z,x_{0},t)\  & =\ \alpha_{DC}(z)\ +\ \alpha_{AC}(z,x_{0})\,\cos(\Omega t)\nonumber \\
\alpha_{DC}(z)\  & =\ \frac{1}{1+i\tilde{z}}\nonumber \\
\alpha_{AC}(z,x_{0})\  & =\ \frac{2ikA_{0}\cos(Kx_{0})\times\exp\left[-B\left(1-i\tilde{z}\right)\right]}{1+i\tilde{z}}
\end{align}

Here, $\tilde{z}=z/z_{R}$, $z_{R}=\pi w_{0}^{2}/\lambda$ is the
Rayleigh range of the focused Gaussian beam, $w_{0}$ its waist radius,
$k=2\pi/\lambda$ the wavenumber, and $B=K^{2}w_{0}^{2}/8$. The complex
collection ratio $\alpha(z,x_{0},t)$ can be separated in two parts:
a static term $\alpha_{DC}(z)$ which is a Lorentzian function of
$z$ and describes the complex collection ratio in the absence of
standing SAWs, and a dynamic term $\alpha_{AC}(z,x_{0})$, which describes
the effect of the time modulation of the light field by the standing
SAWs. This dynamic term not only contains the Lorentzian attenuation
as a function of the defocusing, but also an additional exponential
term that depends on the ratio between the Gaussian waist radius $w_{0}$
and the acoustic wavelength $\Lambda$ via $B$. This exponent $B$
is a complex number, meaning that $\alpha_{AC}(z,t)$ has a certain
periodicity in $z$.

\emph{One-beam experiment.} Here we block the reference arm, so that
the signal is only given by the amplitude modulation. The quantity
responsible for the amplitude modulation is $|\alpha(z,x_{0},t)|^{2}$
which, as shown in Appendix \ref{sec:complex_collection_ratio}, contains
a DC component and components at frequencies $\Omega$ and $2\Omega$.
In our experiment we are only interested in the $\Omega$ component
since the AC coupled RF photodiode blocks the DC component, and the
lock-in amplifier demodulates at $\Omega$. Moreover, the $2\Omega$
term is much smaller in amplitude than the $\Omega$ term, since it
scales like $A_{0}^{2}$, as opposed to $A_{0}$ for the $\Omega$
term. After the demodulation by the lock-in amplifier, we are left
with an rms signal of the amplitude modulation given by:
\begin{align}
V_{amp}^{rms}(z)\  & \propto\ \frac{2|\gamma|e^{-B}}{\sqrt{2}(1+\tilde{z}^{2})}\times\left|\sin(B\tilde{z})\right|,\label{eq:amplitude_modulation}
\end{align}

where $\gamma=2kA_{0}\cos(Kx_{0})$ and the proportionality symbol
means that we are not taking into account the gain provided by photodiode
and lock-in amplifier. This equation gives periodic zeros at positions
that solve the equation $\sin(B\tilde{z})=0$. Substituting $B=K^{2}w_{0}^{2}/8=\pi^{2}w_{0}^{2}/(2\Lambda^{2})$
and $\tilde{z}=z\lambda/(\pi w_{0}^{2})$, we find that the positions
of the zeros is given by $z=n\times z_{T}$, where $n$ is an integer
and $z_{T}=2\Lambda^{2}/\lambda$ is the conventional Talbot length
for a diffraction grating with period $\Lambda$ illuminated by a
plane wave.

\emph{Two-beams experiment.} Here we interfere light from the sample
arm with light from the reference arm. These two light beams are adjusted
to have equal intensities at zero defocusing ($z=0$), meaning $I_{ref}=I_{in}$
in Eq. \ref{eq:Intensity_interference}. We lock the interferometer
at the side of the fringe, such that the total intensity of the interfering
beams is $I_{lock}=2I_{ref}$, which corresponds to the light in the
reference arm having phase $\phi_{ref}=\pm\pi/2$ for $z=0$. If now
we introduce a defocusing ($z\neq0)$, $\alpha_{DC}$ changes both
in amplitude and in phase, lowering the light intensity from the sample
arm. Consequently $\phi_{ref}$ also changes to keep the interferometer
locked. While we can calculate $\phi_{ref}(z)$ from Eq. \ref{eq:Intensity_interference},
imposing that $I_{tot}\ =\ I_{lock}$, a more useful quantity is the
difference between the phase of $\alpha_{DC}(z)$, and the phase of
the reference beam $\Delta\phi(z)=\arg[\alpha_{DC}(z)]-\phi_{ref}(z)$
:

\begin{equation}
\Delta\phi(z)\ \simeq\ \mp\ \mathrm{acos}\left(\frac{\tilde{z}^{2}}{2\sqrt{1+\tilde{z}^{2}}}\right),\label{eq:phase_difference}
\end{equation}

which depends on whether we lock the interferometer on a positive
($-$) or negative ($+$) slope. We note that in the derivation of
this equation we considered $\arg(\alpha_{DC})\simeq\arg(\alpha)_{DC}$,
and $|\alpha_{DC}|\simeq|\alpha|_{DC}$, which is allowed since the
SAW displacement is very small. The rms of the total interferometric
signal is obtained by calculating the rms value of the $\Omega$ component
of Eq. \ref{eq:Intensity_interference}, which for small SAW displacement
$A_{0}$ can be approximated as: 
\begin{align}
V_{tot}^{rms}(z)\  & \simeq\ \frac{2\left|\gamma\right|e^{-B}}{\sqrt{2}}\left|\frac{\sin(\Delta\phi+B\tilde{z})}{\sqrt{1+\tilde{z}^{2}}}+\frac{\sin(B\tilde{z})}{1+\tilde{z}^{2}}\right|.\label{eq:total_interferometric_signal}
\end{align}

The first term in the modulus is from the phase modulation, while
the second term originates from amplitude modulation. In detection,
the phase modulation term can not be separated from the amplitude
modulation term, as it is the result of interference. In mathematical
terms, $V_{tot}^{rms}\neq V_{amp}^{rms}+V_{phase}^{rms}$ due to the
non-linearity of the modulus function.

Now we compare our theory to experiments where we measure both the
amplitude modulation $V_{amp}^{rms}$ and the total interferometric
signal $V_{tot}^{rms}$ as a function of the defocusing distance $z$.
In particular, we show the results corresponding to the situation
where the focused Gaussian beam is centered on an antinode of the
standing SAW, corresponding to $x_{0}=0$ in Eqs. \ref{eq:amplitude_modulation}
and \ref{eq:total_interferometric_signal}. While changing the defocusing
$z$, the relative position $x_{0}$ between the antinode and the
beam spot could change due to imperfect alignment. To compensate for
this, we do a short line scan along the $x$ direction, exemplarily
shown in Fig. \ref{fig:Standing_waves_and_SAW_cavity}, and we average
the amplitudes of the measured peak values.
\begin{figure}[!tph]
\includegraphics[width=1\columnwidth]{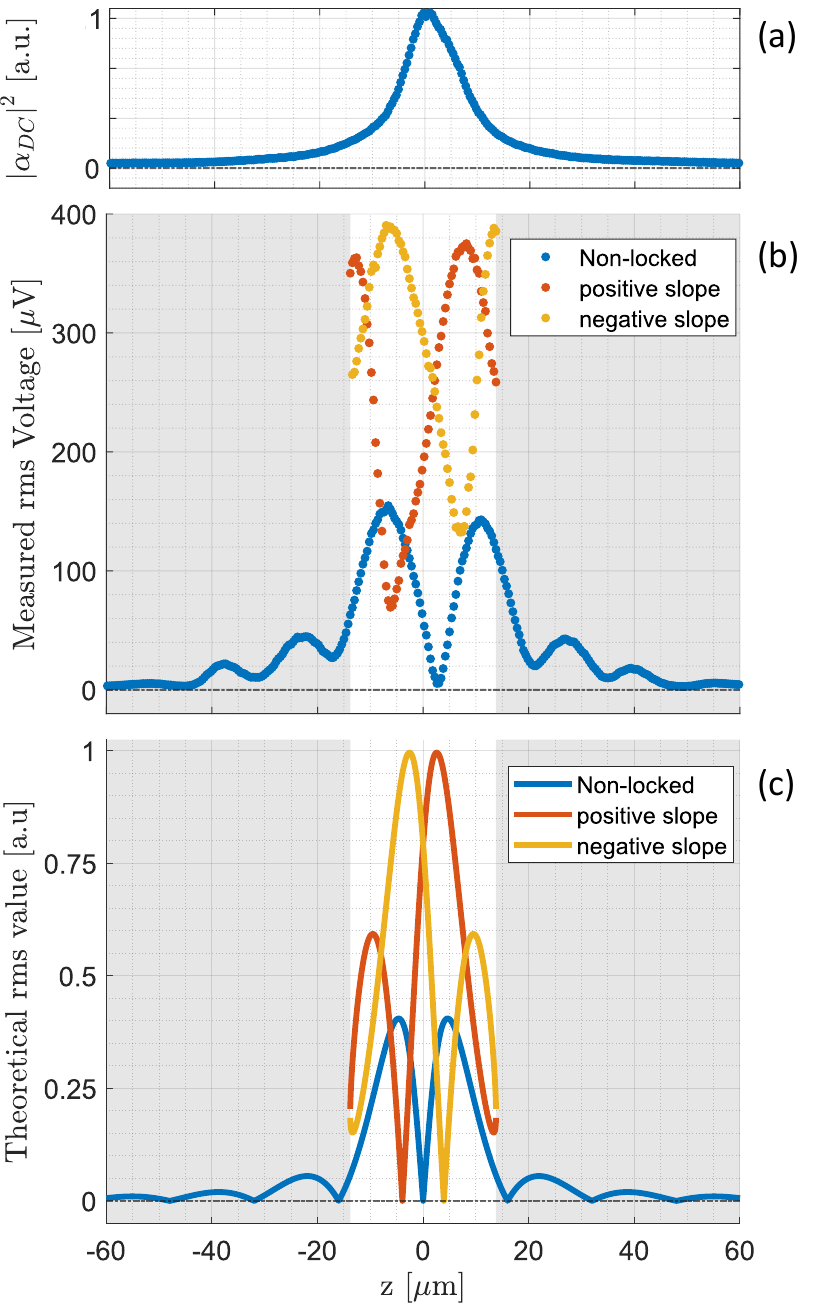}\caption{(a) Experimental DC component of the complex collection ratio as a
function of the defocusing $z$. (b) Experimental and (c) theoretical
$\Omega$ component of the amplitude modulation (blue) and of the
total interferometric signals for the interferometer locked at the
positive (red) or negative (yellow) slope. \protect\label{fig:measurements_amplitude_phase_modulation}}
\end{figure}
 In Fig. \ref{fig:measurements_amplitude_phase_modulation} (a) we
show the measured $|\alpha_{DC}|^{2}$ as a function of $z$, reproducing
the expected Lorentzian dependency. Fig. \ref{fig:measurements_amplitude_phase_modulation}
(b) shows the measured amplitude modulation $V_{amp}^{rms}$ in blue,
as well as the total interferometric signals $V_{tot}^{rms}$, obtained
by locking the interferometer on the positive (red) or negative (yellow)
slope. The measured signal is the rms Voltage detected by the lock-in
amplifier after 1 GHz demodulation. The grey areas are regions where
it is not possible to lock the interferometer due to a decrease in
the light intensity coupled back to the fiber in the sample arm of
the interferometer. Fig. \ref{fig:measurements_amplitude_phase_modulation}
(c) shows the corresponding theoretical data based on Eqs. \ref{eq:amplitude_modulation},
\ref{eq:phase_difference} and \ref{eq:total_interferometric_signal}.

We observe excellent qualitative agreement between measurements and
theory: the signals are almost symmetric with respect to the $z$
defocusing, the total interferometric signals locked at a positive
and negative slope intersect at the defocusing $z$ for which the
amplitude modulation is zero, and the positions of the minima in the
amplitude modulation are close-to-periodic with period $z_{period}=(15.6\pm2.8)\,\text{\ensuremath{\mu}m}$.
This value is close to the expected periodicity from Eq. \ref{eq:amplitude_modulation}:
$z_{T}=2\Lambda^{2}/\lambda=16\,\text{\ensuremath{\mu}m}.$ Residual
misalignment of the sample arm of the interferometer can explain the
few quantitative differences we observe, such as the minimum of the
amplitude modulation located at $z\simeq3\,\text{\ensuremath{\mu}m}$
instead of $z=0$, and the asymmetry in the height of the peaks between
negative and positive defocusing, both in the amplitude modulation
and in the total interferometric signals. Spurious reflections inside
the fiber beam splitter are most likely the reason why the measured
signals do not reach zero, except for the central dip in the amplitude
modulation. 

\begin{figure}[t]
\includegraphics[width=1\columnwidth]{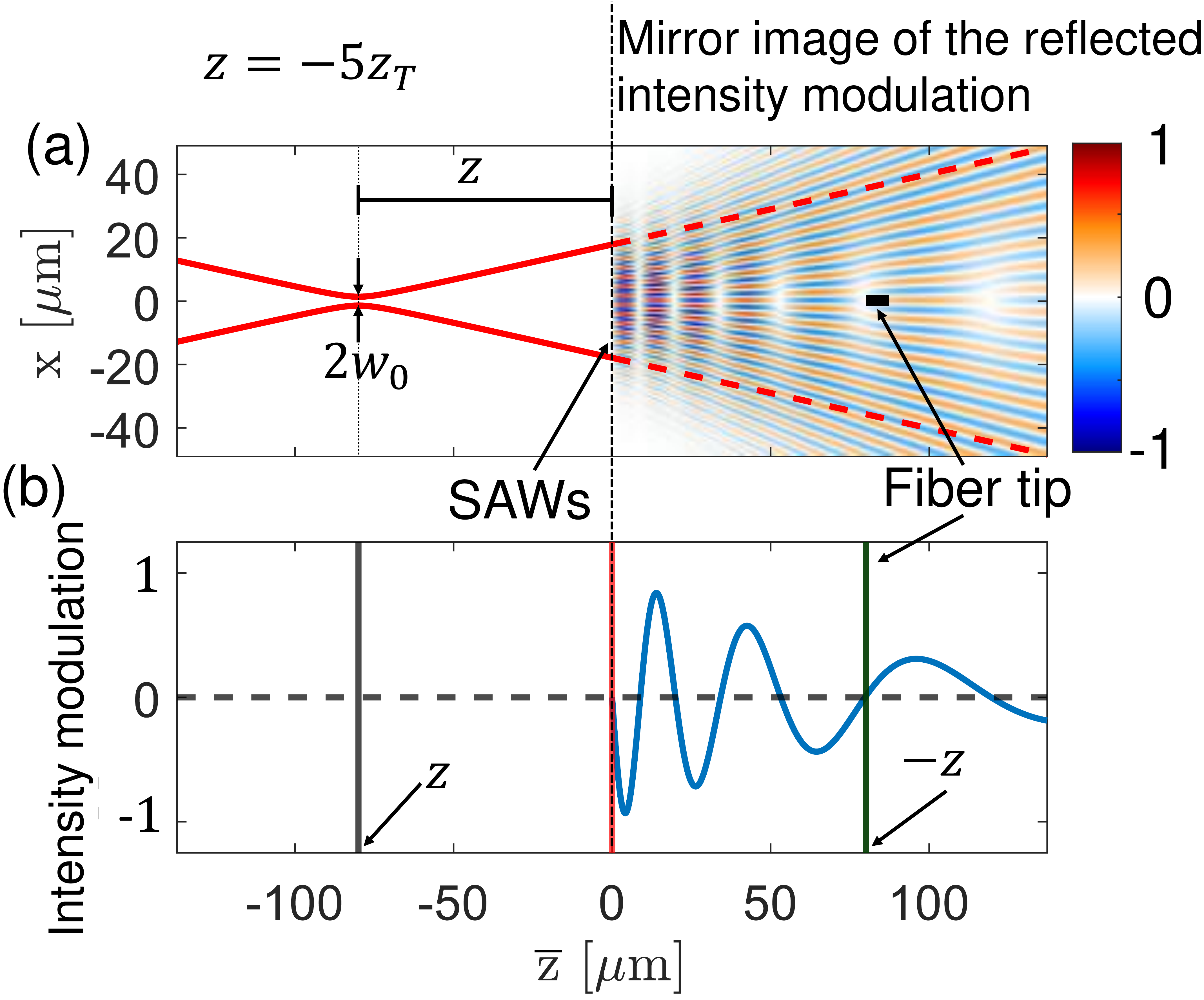}

\caption{\protect\label{fig:Talbot_effect_theory} Example of Talbot pattern
obtained with a beam defocusing of $z=-5z_{T}$ . (a) In the half
plane $\overline{z}<0$, a plot of the Gaussian beam before the grating.
In the half plane $\overline{z}>0$, a plot of the mirrored image
of the reflected light intensity modulation, after subtracting the
constant intensity of the non-diffracted Gaussian beam. We considered
a SAW profile given by Eq. \ref{eq:SSAWs_displacement} , for $x_{0}=0$
and $t=0$. Self-images of the field at the grating appear at specific
positions given by Eq. \ref{eq:non_linear_selfImages_position}, here
visible as zones with zero intensity modulation . (b) Cross-sectional
plot of (a) along $x=0$ showing that, on the optical axis, the intensity
modulation of the reflected field is zero at the positions of the
self-images, and in particular at the mirrored position of the image
of the fiber tip as indicated.}
\end{figure}

\emph{Talbot effect.} We have seen from Eq. \ref{eq:amplitude_modulation}
that the amplitude modulation is periodic in $z$ with the period
$z_{T}=2\Lambda^{2}/\lambda$. To simplify the following math we now
define a new coordinate system $\overline{z}$, with the SAW device
placed at the origin $\overline{z}=0$, and the beam focus is at position
$\overline{z}=z$. For clarity, we unfold the back reflected light
propagation to the right of $\overline{z}=0$, as shown in Fig. \ref{fig:Talbot_effect_theory}.
In this new coordinate system, and due to the symmetric unfolding
with respect to $\overline{z}=0$, the projected image of the fiber
is always at position $\overline{z}=-z$, and the field coupled back
to the fiber after reflection, is the field at this position mode-matched
to the fiber. Since we know the optical field at the grating plane
$\overline{z}=0$, we can propagate it to $\overline{z}=-z$ using
the Fresnel-Kirchhoff diffraction integral, the detailed calculation
is shown in Appendix \ref{sec:Talbot_effect}. In Fig. \ref{fig:Talbot_effect_theory}
(a) we show the intensity of the propagated light field after reflection
from the grating. Contrary to the case of a grating illuminated by
a plane wave, under Gaussian beam illumination, the Talbot self-images
are laterally magnified by a factor $M_{G}=R_{\overline{z}}/(R_{\overline{z}}-\overline{z})$
\citep{szapielFresnelDiffraction1979}, where $R_{\overline{z}}=(\overline{z}-z)\times(1+z_{R}^{2}/(\overline{z}-z)^{2})$
is the radius of curvature of the Gaussian beam at a generic observation
plane $\overline{z}$. Also the positions of the self images are non-periodic
in $\overline{z}$ and given by solutions to \citep{szapielFresnelDiffraction1979} 

\begin{equation}
m\frac{z_{T}}{2}\ =\ \frac{R_{\overline{z}}-\overline{z}}{R_{\overline{z}}}\times\overline{z},\label{eq:non_linear_selfImages_position}
\end{equation}
where $m$ is a positive integer. 

Combining all this, we find that, remarkably, when the beam focus
is at a distance $z=n\times z_{T}$ from the grating, with $n$ an
integer number, there is always one of the magnified Talbot self-images
at the position $\overline{z}=-z$, coinciding with the position of
the the fiber and shown for $n=5$ in Fig. \ref{fig:Talbot_effect_theory}
(a). Again, because the surface motion does not give a $\Omega$ signal
if it is in focus, also at these Talbot replica positions, the amplitude
modulation signal vanishes (see also Appendix \ref{sec:Talbot_effect}).

In Fig. \ref{fig:Talbot_effect_theory} (a), we can also observe that
the region of constant intensities are concentrated close to the optical
axis ($x=0$) - this is because we are illuminating only a small portion
of the grating. We note that the number of self-images obtained after
reflection depends on the Gaussian beam radius at the grating, and
therefore on the number of grating periods that are illuminated. In
particular, the minimum beam radius needed to observe a number $m$
of self-images is given by (see Appendix \ref{sec:Talbot_effect}):

\begin{equation}
w_{min}(m)\ =\ \Lambda\sqrt{\frac{2m}{\pi}}.\label{eq:number_of_selfImages}
\end{equation}

\emph{Conclusions.} By using a fiber-based Michelson interferometer
to measure GHz standing SAWs in a SAW device, we discovered that on
top of the usual phase modulation term, the total interferometric
signal also has a significant amplitude modulation term. We showed
that the amplitude modulation vanishes not only if the device is in
focus, but also at periodic defocusings from the SAW device, corresponding
to multiples of the Talbot length. This is explained by a combined
change of the radius of curvature and the Talbot effect for curved
wave fronts \citep{szapielFresnelDiffraction1979}. Our findings also
show that an interferometric setup is not always the optimal solution
for measuring spatially-resolved oscillating surface displacements.
Simple amplitude measurements with a single single-mode fiber can
be an easier way and the fiber splitter can be replaced by a fiber
circulator to increase the signal to noise ratio. Our theory shows
that the signal corresponding to the amplitude modulation is even
stronger than the interferometric signal if the SAW period $\Lambda$
is smaller than the beam waist: $\Lambda\lesssim w_{0}$, as shown
in Appendix \ref{sec:detection_with_fiber_circulator}.

\textbf{Acknowledgements.} We acknowledge funding from a NWO Vrij
Programma Grant (QUAKE, 680.92.18.04), the European Union’s Horizon
2020 research and innovation programme under grant agreement No. 862035
(QLUSTER), and the Quantum Software Consortium.

\bibliographystyle{apsrev4-1}
\bibliography{Bibliography_5}

\appendix

\section{the complex collection ratio}

\label{sec:complex_collection_ratio}Here we derive the complex collection
ratio $\alpha(z,x_{0},t)$. We use the shifted reference frame where
the focus of the Gaussian beam, as projected image of the input fiber,
is the origin of the optical axis, and the reflecting surface is at
a distance z. This shift does not change the final results that depend
only on the relative distance between the SAW device and the focus
of the Gaussian beam. We consider the ideal case where the imaging
system produces a perfect image of the fiber tip in the focus. This
field can be written as $E_{in}(x,y,z,t)\ =\ \psi_{in}^{+}(x,y,z)\ \exp(-ikz+i\omega t)$
with a forward-propagating Gaussian mode (without Gouy phase)

\begin{align}
\psi_{in}^{+}(x,y,z)\  & =\ \frac{\sqrt{2}}{\sqrt{\pi}w(z)}\times\exp\left[-ik\frac{x^{2}+y^{2}}{2q(z)}\right]\ \\
 & =\ \frac{\sqrt{2}}{\sqrt{\pi}w(z)}\times\exp\left[-\frac{x^{2}+y^{2}}{w^{2}(z)}(1+i\tilde{z})\right],\nonumber 
\end{align}

where $q(z)=z+iz_{R}$ is the complex beam parameter, $z_{R}=\pi w_{0}^{2}/\lambda$
is the Rayleigh range, $w_{0}$ is the beam waist radius located at
$z=0$, $w(z)=w_{0}\sqrt{1+\tilde{z}^{2}}$ is the beam radius at
position $z$, and $\tilde{z}=z/z_{R}$. The field is normalized:
\begin{equation}
\iint|\psi_{in}^{+}(x,y,z)|^{2}\ dx\ dy\ =\ 1.
\end{equation}

The complex collection ratio defined in the main text can be calculated
via the overlap integral between the reflected backward-propagating
field, and the backward-propagating image of the input field $\psi_{in}^{-}=(\psi_{in}^{+})^{*}.$
This overlap integral can be calculated at any $z$ plane, since optical
propagation is a unitary operation. We calculate it at the plane of
the SAW device. For a flat and perfectly reflecting interface, the
reflected field mimics the input field $E_{refl}^{-}=E_{in}$, and
we obtain the DC coupling: 

\begin{align*}
\alpha_{DC}\  & =\ e^{2ikz}\iint\psi_{in}^{+}(x,y,z)^{2}\ dx\ dy\ =\ \frac{e^{2ikz}}{1+i\tilde{z}}.
\end{align*}

We can calculate the in-coupling efficiency $\eta$ as 

\begin{equation}
\eta\ =\ |\alpha_{DC}|^{2}\ =\ \frac{I_{col}}{I_{in}}\ =\ \frac{1}{1+\tilde{z}^{2}},
\end{equation}
which corresponds to the DC coupling when we displace the reflecting
surface by an amount $z$, with respect to the position of the beam
waist.

We now consider how standing surface acoustic waves (SAWs) modulate
the complex collection ratio $\alpha$. In presence of standing SAWs,
the reflecting surface oscillates with profile

\begin{equation}
\Delta z(x,y)\ =\ A(t)\ \cos[K(x-x_{0})],\label{eq:delta_displacement}
\end{equation}

where $K=2\pi/\Lambda$ is the SAW wave number, $\Lambda$ is the
SAW wavelength, $A(t)=A_{0}\ \cos(\Omega t)$ is the displacement
at an anti-node with amplitude $A_{0}$, $\Omega$ is the SAW angular
frequency, and $x_{0}$ indicates the position of the laser beam.
The reflected field becomes 

\begin{equation}
E_{refl}^{-}=\exp(2ik\Delta z)E_{in},\label{eq:reflected_field}
\end{equation}
where $\Delta z>0$ is away from the input fiber. From Eq. \ref{eq:reflected_field},
we remove the stationary phase factor $\exp(2ikz)$, and we obtain
\begin{align}
\alpha(z,x_{0},t) & =\iint\exp(2ik\Delta z)\psi^{+}(x,y,z)^{2}\ dx\ dy.\label{eq:alpha_generic}
\end{align}

Substituting \ref{eq:delta_displacement} into \ref{eq:alpha_generic},
we arrive at the following integral:

\begin{align}
\alpha(z,x_{0},t)\  & =\ \iint\exp\{2ikA(t)\cos[K(x-x_{0})]\}\\
 & \times\ \exp\left[-\frac{2(x^{2}+y^{2})(1+i\tilde{z})}{w^{2}(z)}\right]\ \frac{dx\ dy}{\pi w^{2}(z)}.\nonumber 
\end{align}

This integral can be solved by separating the integral in $x$ and
$y$ direction:

\begin{align}
\alpha(z,x_{0},t)\  & =\ \frac{2}{\pi w^{2}(z)}\ I_{y}\ I_{x},\ \text{where}\label{eq:alpha_SAW}\\
I_{y}\  & =\ \int_{-\infty}^{+\infty}\exp\left[-\frac{2y^{2}}{w^{2}(z)}(1+i\tilde{z})\right]\ dy\nonumber \\
I_{x}\  & =\ \int_{-\infty}^{+\infty}\exp\left[-\frac{2x^{2}}{w^{2}(z)}(1+i\tilde{z})\right]\ \times e^{2ik\Delta z}\ dx.\nonumber 
\end{align}

$I_{x}$ and $I_{y}$ can be solved by using the standard integral

\begin{equation}
\int_{-\infty}^{+\infty}\exp\left[-ax^{2}+ibx\right]\,dx\ =\ \sqrt{\frac{\pi}{a}}\exp\left(-\frac{b^{2}}{4a}\right).\label{eq:standard_gaussian_integral}
\end{equation}

While evaluation of $I_{y}$ is straightforward, yielding

\begin{equation}
I_{y}=\sqrt{\frac{\pi}{2(1+i\tilde{z})}}w(z),
\end{equation}

evaluating $I_{x}$ requires a few more steps: first we separate the
generic standing SAW displacement in Eq. \ref{eq:delta_displacement}
into the sine and cosine quadratures:

\begin{equation}
\cos\left[K(x-x_{0})\right]\ =\ \cos(Kx)\cos(Kx_{0})+\sin(Kx)\sin(Kx_{0})\ ,
\end{equation}

then, since $kA(t)<<1$, we can expand the exponential:
\begin{flushleft}
\begin{multline}
\exp\{2ikA(t)\cos[K(x-x_{0})]\simeq\\
1+2ikA(t)[\cos(Kx)\cos(Kx_{0})+\sin(Kx)\sin(Kx_{0})].\label{eq:exponential_approximated}
\end{multline}
\par\end{flushleft}

Multiplication of this term by the Gaussian function, and their integration,
leads to $I_{x}$. Since $\sin(Kx)$ is an odd function of $x$ whereas
the Gaussian function is even, one integral vanishes and we obtain:

\begin{align}
I_{x} & =I_{y}+2ikA(t)\cos(Kx_{0})\\
 & \times\int_{-\infty}^{+\infty}\exp\left[-\frac{2x^{2}}{w^{2}(z)}(1+i\tilde{z})\right]\cos(Kx)\,dx.\nonumber 
\end{align}

We can use the expansion $\cos[Kx]=[\exp(iKx)+\exp(-iKx)]/2$ to obtain
the following equation:

\begin{align}
I_{x} & =I_{y}+ikA(t)\cos(Kx_{0})\times(G^{+}+G^{-}),\ \mathrm{where}\label{eq:x_integral}\\
G^{\pm} & =\int_{-\infty}^{+\infty}\exp\left[-\frac{2x^{2}}{w^{2}(z)}(1+i\tilde{z})\right]\times\exp(\pm iKx)\ dx.\nonumber 
\end{align}

The physical meaning of this equation is the following: for a small
SAW displacement $A_{0}$, the reflecting surface behaves as an amplitude
diffraction grating with cosine profile. The Gaussian beam impinges
on this diffraction grating, and the back reflection is coupled back
to the fiber (term $I_{y}$ in Eq. \ref{eq:x_integral}) . The diffraction
grating also generates two tilted Gaussian beams with angles $\theta^{\pm}=\pm\lambda/\Lambda$,
corresponding to the terms $G^{+}$ and $G^{-}$ in Eq.\ref{eq:x_integral},
as will be shown in Appendix \ref{sec:Talbot_effect}. By using once
again the standard integral \ref{eq:standard_gaussian_integral},
we obtain:

\begin{equation}
G^{\pm}=\sqrt{\frac{\pi}{2(1+i\tilde{z})}}w(z)\exp\left[-B(1-i\tilde{z})\right],\label{eq:G_plus_minus}
\end{equation}

where $B=-K^{2}w_{0}^{2}/8$. Inserting Eq.\ref{eq:G_plus_minus}
into $I_{x}$, and $I_{x}$ into Eq. \ref{eq:alpha_SAW}, we get:

\begin{equation}
\alpha(z,x_{0},t)=\frac{1}{1+i\tilde{z}}\left\{ 1+2ikA(t)\cos(Kx_{0})\exp\left[-B(1-i\tilde{z})\right]\right\} \ .
\end{equation}

The amplitude modulation term is given by $|\alpha(z,t)|^{2}$, which
after some calculations can be expressed as:

\begin{align}
|\alpha|^{2}\  & =\ |\alpha|_{DC}^{2}\ +\ |\alpha|_{\Omega}^{2}\cos(\Omega t)\ +\ |\alpha|_{2\Omega}^{2}\cos(2\Omega t)\\
|\alpha|_{DC}^{2}\  & =\ \frac{1}{1+\tilde{z}^{2}}\nonumber \\
|\alpha|_{\Omega}^{2}\  & =\ -\frac{2\gamma\times\exp\left(-B\right)\times\sin\left(B\tilde{z}\right)}{1+\tilde{z}^{2}}\nonumber \\
|\alpha|_{2\Omega}^{2}\  & =\ \frac{\gamma^{2}\times\exp\left(-2B\right)}{1+\tilde{z}^{2}},\nonumber 
\end{align}

where $\gamma=2kA_{0}\cos(Kx_{0})$. 

\renewcommand{\theequation}{B\arabic{equation}}\setcounter{equation}{0} 

\section{The Talbot effect}

\label{sec:Talbot_effect}Here we explicitly calculate the propagated
field after the grating, used to create the plots in Figs. \ref{fig:Sketch_of_experiment}
and \ref{fig:Talbot_effect_theory} that visualize the Talbot patterns
created by diffraction of a Gaussian beam from a periodic grating.
Since the grating affects the propagated field only along the $x$
transversal direction, for simplicity we study the evolution of a
1D Gaussian beam after reflection from the surface. On the optical
axis, we define a new coordinate system $\overline{z}$ with the origin
at the surface of the SAW device. In this coordinate system, the beam
focus is at position $\overline{z}=z$, where $z$ is the beam defocusing
introduced in the calculations for the complex collection ratio. For
clarity, and as illustrated in Fig. \ref{fig:Talbot_effect_theory}
in the main text, we unfold the reflected field to the right of $\overline{z}=0$.
The field at the surface ($\overline{z}=0$) is given by the function

\begin{equation}
\psi_{reflect}^{-}(\overline{z}=0,x,t)=\psi_{in}^{+}(z)\times\exp\left[2ikA(t)\cos(Kx_{0})\cos(Kx)\right],\label{eq:field_at_surface}
\end{equation}
where we omitted the sine quadrature in the expansion of the displacement
(Eqs. \ref{eq:delta_displacement}, \ref{eq:exponential_approximated})
due to integration in the next step. The field $g(x,\overline{z},t)$
at a generic position $\overline{z}$ is given by convoluting the
input field $\psi_{reflect}^{-}(\overline{z}=0,x,t)$ with the impulse
response function of free space in the Fresnel approximation (Fresnel-Kirchhoff
diffraction integral):

\begin{equation}
h(x,\overline{z})\simeq h_{0}\exp\left(-i\frac{\pi}{\lambda\overline{z}}x^{2}\right),
\end{equation}

where $h_{0}=i/(\lambda\overline{z})$, leading to

\begin{equation}
g(x,\overline{z},t)=h_{0}\int_{-\infty}^{+\infty}\psi_{reflect}^{-}(x',t)\times\exp\left[-i\frac{\pi}{\lambda\overline{z}}\left(x-x'\right)^{2}\right]\ dx'.\label{eq:output_field}
\end{equation}

Following the same steps we used to calculate the complex collection
ratio $\alpha(z,t)$, we expand the exponential containing the cosine
term:

\begin{align}
\exp\left[i\gamma(t)\cos(Kx)\right] & \simeq1+i\gamma(t)\cos(Kx),\\
\mathrm{where\ }\gamma(t)=2kA_{0}\cos(Kx_{0})\cos(\Omega t).\nonumber 
\end{align}

We can expand the cosine in exponential form: $\cos(Kx)=\left[\exp(iKx)+\exp(-iKx)\right]/2$,
and finally we can rewrite Eq. \ref{eq:output_field} as

\begin{align}
g(x,\overline{z},t) & =\int_{-\infty}^{+\infty}\psi_{in}^{+}\times h(x-x',\overline{z})\ dx'\label{eq:propagated=000020field}\\
 & +\frac{i\gamma(t)}{2}\int_{-\infty}^{+\infty}\psi_{in}^{+}\times e^{iKx'}\times h(x-x',\overline{z})\ dx'\nonumber \\
 & +\frac{i\gamma(t)}{2}\int_{-\infty}^{+\infty}\psi_{in}^{+}\times e^{-iKx'}\times h(x-x',\overline{z})\ dx'.\nonumber 
\end{align}

The first term describes the evolution of a Gaussian beam, and therefore
represents the back-reflected Gaussian beam, as if it was reflected
from a flat surface. The second and third terms describe two tilted
beams at an angle of $\theta=\pm\lambda/\Lambda$ with respect to
the optical axis, as we can see by writing the second (+) and third
(-) terms as:

\begin{align}
g_{\pm}=\frac{i\gamma(t)h_{0}C}{2}\exp\left(-\frac{i\pi x^{2}}{\lambda\overline{z}}\right)\int_{-\infty}^{+\infty}\exp\left[-ax'{}^{2}\right]\\
\times\exp\left[ix'\frac{2\pi}{\lambda\overline{z}}\left(x\pm\overline{z}\frac{\lambda}{\Lambda}\right)\right]\ dx' & ,\nonumber 
\end{align}

where 
\begin{equation}
a=\left(\frac{1+iz/z_{R}}{2w_{0}^{2}(1-z^{2}/z_{R}^{2})}+\frac{i\pi}{\lambda\overline{z}}\right),
\end{equation}
and $C$ is a normalization factor. The integral in Eq. \ref{eq:propagated=000020field}
can be solved by using the standard integral from Eq. \ref{eq:standard_gaussian_integral}.
We do not show here the lengthy expression, but an example solution
is shown in Fig. \ref{fig:Talbot_effect_theory} in the main text,
where we evaluated the integral at $t=0$ and $z_{0}=5z_{T}$. As
a result of the interference between the diffracted beams, self-images
of the field at the grating appear at non-periodic positions on the
optical axis. These positions are dependent on the radius of curvature
of the field impinging on the grating, and therefore on the beam defocusing
$z$. Therefore different values of $z$, leads to different Talbot
patterns.

\subsection{Existence and positions of the self images}

Here we solve Eq. \ref{eq:non_linear_selfImages_position} to show
that for beam defocusings $z=nz_{T}$, where $n$ is an integer, one
of the self-images is always located at $\overline{z}\simeq-z$. In
our coordinate system, with the unfolded reflection, this corresponds
to the position of the projection of the image of the fiber. We will
also give a proof for Eq. \ref{eq:number_of_selfImages}. 

Starting by substitution of $R_{\overline{z}}=(\overline{z}-z)(1+z_{R}^{2}/(\overline{z}-z)^{2})$
into Eq. \ref{eq:non_linear_selfImages_position}, we obtain a second
order equation $a\overline{z}^{2}+b\overline{z}+c=0,$where $a=z+mz_{T}/2$,
$b=-(mzz_{T}+z^{2}+z_{R}^{2})$, $c=mz_{T}(z^{2}+z_{R}^{2})/2$. This
equation has two solutions given by

\begin{equation}
\overline{z}_{1,2}\ =\ \frac{m\ z\ z_{T}+z^{2}+z_{R}^{2}\pm(z^{2}+z_{R}^{2})\sqrt{1-\delta^{2}}}{2(z+mz_{T}/2)},\label{eq:generic_solution_self_images}
\end{equation}

where $\delta\ =\ mz_{T}z_{R}/(z^{2}+z_{R}^{2})$. The first thing
we notice is that the solutions only exist if $\delta^{2}\leq1$,
which, under the condition that $m>0$, leads to the equation:
\begin{equation}
z_{0}^{2}\ \geq\ z_{R}\ (m\ z_{T}-z_{R}),\label{eq:number_of_self_images_intermediate}
\end{equation}

which gives the minimum distance between the beam waist of the focused
Gaussian beam and the grating, in order to observe a number $m$ self
images. This equation can be rewritten in terms of the beam radius
at the grating, showing that in order to see $m$ self-images, the
minimum beam radius at the grating has to be 
\begin{equation}
w_{min}(m)\ =\ \Lambda\ \sqrt{\frac{2m}{\pi}}.
\end{equation}

We can now see what happens to Eq. \ref{eq:generic_solution_self_images}
when $z\ =\ nz_{T}$, in the limit that $\delta\ll1$, where we can
approximate $\sqrt{1-\delta^{2}}\simeq1$:
\begin{equation}
\overline{z}_{1,2}\ \simeq\frac{m\ n\ z_{T}^{2}+n^{2}z_{T}^{2}+z_{R}^{2}\pm(n^{2}z_{T}^{2}+z_{R}^{2})}{z_{T}(2n+m)},\label{eq:approximated_solution_self_images}
\end{equation}

and by choosing the solution with the (-) sign, we obtain $\overline{z}_{1}\ \simeq z_{T}\times mn/(2n+m)$,
and for $n=-m$ we get $\overline{z}_{1}=-nz_{T}$, proving that the
$n$-th self image lies at the position $\overline{z}\simeq-z.$ To
check if $\delta<<1$, we can introduce a variable $a$ equal to the
ratio between the SAW wavelength and the beam waist radius, such that
$\Lambda=aw_{0}$, and we can write
\begin{equation}
\delta\ =\ \frac{2\pi\ m\ a^{2}}{\pi^{2}+4\ m^{2}a^{4}}.
\end{equation}

We can see that $\delta\sim m^{-1}$ when $m\to\infty$, therefore
the approximation is better for high values of $m$. In our experiment
$a=2$ ($\Lambda=2.8\ \mu m$ and $w_{0}=1.4\ \mu m$), and already
for $m=1$ the approximation used in Eq. \ref{eq:approximated_solution_self_images}
is valid.

\renewcommand{\theequation}{C\arabic{equation}}\setcounter{equation}{0} 

\section{SAW detection with a fiber circulator}

\label{sec:detection_with_fiber_circulator}
\begin{figure}[H]
\includegraphics[width=1\columnwidth]{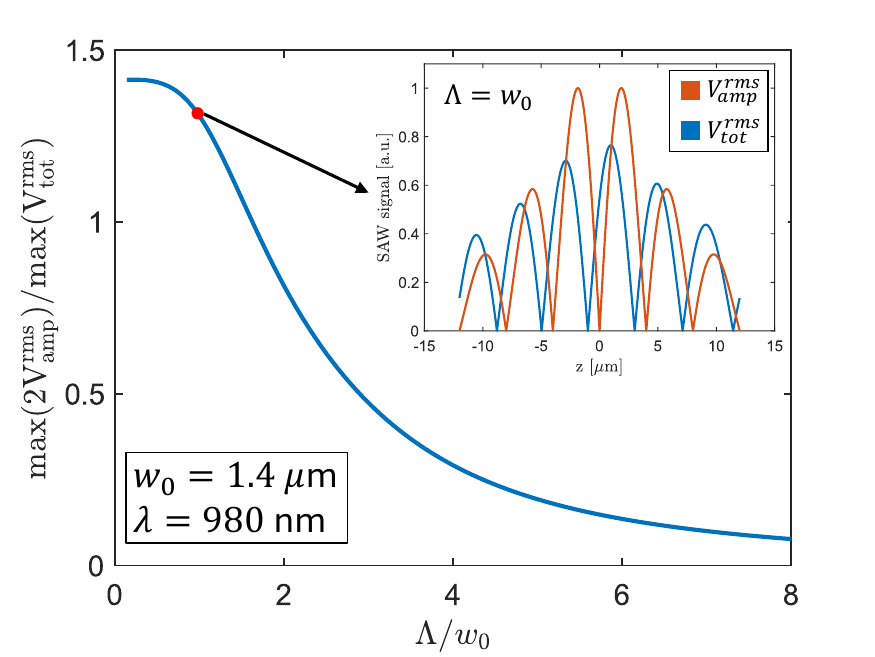}

\caption{\protect\label{fig:fiber_circulator_detection}Ratio $r=\mathrm{max[2V_{amp}^{rms}(z)]/max[V_{tot}^{rms}(z)]}$,
as a function of the acoustic wave $\Lambda$. The plot has been calculated
from Eqs. \ref{eq:amplitude_modulation} and \ref{eq:total_interferometric_signal},
with $w_{0}=1.4\,\mathrm{\mu m}$ and $\lambda=980\,\mathrm{nm}$.
The inset shows the two signals $2V_{amp}^{rms}(z)$ (red) and $V_{tot}^{rms}(z)$
(blue) for $\Lambda/w_{0}=1$.}
\end{figure}

An important result of
Eqs. \ref{eq:amplitude_modulation} and \ref{eq:total_interferometric_signal}
is that, under some circumstances, SAW displacements can be better
measured without an interferometer. The amplitude modulation signal
can even be larger than the total interferometric signal if we exchange
the fiber splitter in Fig. \ref{fig:experimental_setup} with a fiber
circulator. In this way the optical power is not split in two, and
the amplitude modulation signal is twice as big as the amplitude modulation
signal present in the normal interferometric setup. To see this we
choose $w_{0}=1.4\,\mathrm{\mu m}$ and $\lambda=980\,\mathrm{nm}$,
and we calculate the ratio $r=\mathrm{max[2V_{amp}^{rms}(z)]/max[V_{tot}^{rms}(z)]}$,
where the factor 2 takes into account the extra optical power available
with the fiber circulator. The result is shown in Fig. \ref{fig:fiber_circulator_detection},
where it is visible that for $\Lambda/w_{0}\lesssim1.6$, measuring
with the fiber circulator leads to a stronger detected signal. 

\end{document}